\documentclass{ws-ijmpa}
\usepackage[super,compress]{cite}
\usepackage{graphicx}
\begin{document}
\markboth{Sumit Panganti and Siba Prasad Das}{A short introduction on Angular momentum of Kerr Blackhole}

%
\catchline{}{}{}{}{}
%

\title{A short introduction on Angular momentum of Kerr Blackhole}

\author{Sumit Panganti and Siba Prasad Das}

\address{Department of Physics, Shivaji University, 
Kolhapur-416004, Maharashtra, India\\
spd.phy@unishivaji.ac.in} 

\maketitle

\begin{history}
\received{22 May 2024}
\revised{22 May 2024}
\end{history}

\begin{abstract}
General relativity (GR) predicts the existence of black hole (BH). The rotating BH called as 
a Kerr Black hole and GR implies that there is an upper limit on the angular momentum per 
mass squared of black holes $\leq 1$, above which the event horizon of the Kerr BH is 
not exist. We find the radial equation for equatorial motion for Kerr BH in terms of 
the effective potential. We have shown the effective potential profile for different 
rotation parameter ($a$). We find the solution of the radial equation of the Kerr  
metric and found the expression of the angular momentum per unit mass squared, $\tilde a = \frac{a}{M}$. 
We showed the profile of $\tilde a$ as a function of $\frac{r}{M}$. The solution also leads 
the energy per unit rest mass ($e$) and we showed its behavior as a function of $\frac{r}{M}$. 
We enumerated the maximum values of radius of innermost stable circular orbit ($r_{ISCO}$) for 
$\tilde a=1$.
\end{abstract}

\ccode{PACS numbers: 04.70.Bw}


\section{Introduction}

The horizon-scale images of black holes (BHs) and their shadows have opened an unprecedented 
window onto tests of gravity~\cite{Vagnozzi:2022moj}. The BH naturally emerges from 
the theory of Gravity called as General Relativity~\cite{hartle,weinberg}. The Kerr metric \cite{kerr} describes 
the geometry of empty space-time around a rotating uncharged axially symmetric black hole (BH) 
with a quasi-spherical event horizon. The Kerr metric is an exact 
solution of the Einstein field equations of general relativity. A distinctive 
prediction of general relativity for the Kerr metric is that a rotating body should exhibit frame-dragging 
known as Lense–Thirring precession. In the case of Kerr BH at close enough distances, 
all objects even light must rotate with the BH; the region where this holds is called
the ergosphere. The Kerr BH have surfaces where the metric seems to have apparent singularities; 
the size and shape of these surfaces depends on the BH's mass and angular momentum. 
The outer surface encloses the ergosphere and has a shape similar to a flattened sphere. 
The inner surface marks the event horizon -- objects passing into the interior of this 
horizon can never again communicate with the world outside that horizon. Objects between 
these two surfaces must corotate with the rotating BH. This feature can 
in principle be used to extract energy, from a rotating BH, equal to its 
invariant mass energy.

In some extension of Einstein gravity the Kerr black hole has been studied recently 
in \cite{Donmez:2023kmh,Donmez:2022dze}. Nearly static magnetized Kerr black-hole in 
nonlinear electrodynamics has been studied recently in \cite{Managave:2023rhn,Redekar:2023qra}. 

In this paper we found the radial equation for equatorial motion for Kerr BH in terms of 
the effective potential. We found the solution of the radial equation of the Kerr  
metric and found the expression of the angular momentum per unit mass squared, $\tilde a = \frac{a}{M}$. 
We showed the variations of $\tilde a$ as a function of $\frac{r}{M}$. Finally we 
enumerated the maximum values of radius of innermost stable circular orbit ($r_{ISCO}$).

The paper is organized is as follows. In Sec.2 we studied the Kerr geometry and the 
effective potential. In Sec.3 we have shown the solutions to the radial equation of 
the Kerr Metric. The numerical studies and correlation among various parameters have 
been analyzed in Sec.4. We conclude in Sec.5.

\section {Kerr Geometry}

For the Kerr BH with mass $M$, angular momentum $J$ and the coordinates$(t,r,\theta,\phi)$ the 
line element of space-time (in geometrized units with $G = c = 1$) is given by\cite{kerr,bardeen}:
\begin{equation}\label{eqnkerr}
{\displaystyle ds^2=-(1-\frac{2Mr}{\rho^2})dt^2-\frac{4Mar\sin^2{\theta}}{\phi^2}d\phi dt+\frac{\rho^2}{\Delta}dr^2+\rho^2d\theta^2+\newline (r^2+a^2+\frac{2Mra^2\sin^2{\theta}}{\rho^2})\sin^2{\theta}d\phi^2}
\end{equation}
where,\\
$\rho^2=r^2+a^2\cos^2{\theta}$ ,\\
$\Delta=r^2-2Mr+a^2$ \\
$a=\frac{J}{M}$\\
It is obvious from the line-element that the corresponding tensor is not the diagonal form 
due to the presence $d\phi dt$ term. We see that the tensor corresponding to the 
line element in Eq.\ref{eqnkerr} is given by:
\begin{equation}\label{tensorkerr}
{\displaystyle g_{\mu\nu}=\begin{pmatrix}
-(1-\frac{2Mr}{\rho^2}) & 0 & 0 & -\frac{2Mar\sin^2\theta}{\rho^2}\\
0 & \frac{\rho^2}{\Delta} & 0 & 0\\
0 & 0 & \rho^2 & 0\\
 -\frac{2Mar\sin^2\theta}{\rho^2} & 0 & 0 & (r^2+a^2+\frac{2Ma^2r\sin^2\theta}{\rho^2} )\sin^2{\theta}
\end{pmatrix}}
\end{equation}
with off-diagonal elements $g_{t\phi} = g_{\phi t}$. The metric is a solution 
of the vacuum Einstein equation and we shall explore some of its most important properties.

It follows from the Kerr metric that it is independent of $t$ and hence stationary.
The metric is also independent of $\phi$ thus there exist Killing vectors $\xi^\mu=(1,0,0,0)^\tau$ 
and $\eta^\mu = (0,0,0,1)^\tau$. 
The Eq.\ref{eqnkerr} reveals singularities when $\rho$ and $\Delta$ vanishes. 
The singularity for $ \rho=0$  arises when $r=0 $ and  $\theta=\frac{\pi}{2}$, 
which corresponds to a real, physical singularity and is the generalization 
of the physical singularity in the Schwarzschild metric~\cite{bardeen}.

We have two solutions for $\Delta=0$:

\begin{equation}\label{kerrradius}
{\displaystyle r\pm=M\pm\left(M^2-a^2 \right)^{\frac{1}{2}}}
\end{equation}

The event horizon does not exist for $a> M$ (as this is a coordinate singularity for Schwarzschild 
metric). Furthermore it is clear that under the condition $a = 0$, Eq.\ref{kerrradius} 
reduces to $r_+ = 2M$ (i.e., event horizon in the Schwarzschild geometry). Thus Eq.\ref{kerrradius} 
defines the generalization of the event horizon. Evidently then the event horizon ranges 
from $ r = 2M$ (when $ a = 0$) to $r=M$ (when $a=M$ for extremal black holes). It can be saying 
that as the angular momentum per mass of black holes increases, the radius of the event horizon 
gets smaller. From Eq.\ref{kerrradius} we also find that the Kerr parameter $a\leq M$, 
a limitation which does not apply to other astronomical objects such as stars. 

In general, the orbits of particles in the Kerr geometry are not confined to a plane unlike in 
Schwarzschild geometry, due to absence of spherical symmetry. However the metric is still 
asymmetric so there exist orbits in the equatorial plane, and such orbits are especially 
interesting since the accretion disk lies close to the equatorial plane\cite{carter}. We will restrict 
our attention to such orbits and for the equatorial plane we then have  $u^\theta = 0$. 

Following Eq.\ref{eqnkerr} the Kerr geometry reduces to:

\begin{equation}\label{eqnkerrnew}
{\displaystyle  ds^2=-\left(1-\frac{2M}{r}\right)dt^2-\frac{4Ma}{r}d\phi dt+\frac{r^2}{\Delta}dr^2+\left(r^2+a^2+\frac{2Ma^2}{r}\right)d\phi^2}.
\end{equation}

Since the Kerr metric is independent of $t$ and $\phi$ there exist {\it two} conserved quantities.
\begin{equation}\label{eqnkerrconserved}
{\displaystyle -e=g_{\mu\nu}\xi^\mu u^\nu ~~~~~~and~~~~~~l=g_{\mu\nu} \eta^\mu u^\nu}
\end{equation}
where $e$ is the energy per unit rest mass and $l$ is the angular momentum per unit rest mass. 
The $\xi^\mu$ and $\eta^\mu$ are two Killing vectors. 
From the Kerr metric it is clear that both the quantities in Eq.\ref{eqnkerrconserved} are 
linear combinations of $u^t$ and $u^\phi$ according to 
\begin{equation}\label{eqnkerrconservednew1}
{\displaystyle -e=g_{tt}u^t+g_{t\phi} u^\phi }
\end{equation}
\begin{equation}\label{eqnkerrconservednew2}
{\displaystyle l=g_{\phi t}u^t+g_{\phi\phi} u^\phi }
\end{equation}

One can invert to solve $u^t$ and $u^\phi$ and are as follows:
\begin{equation}\label{eqncon1}
{\displaystyle u^t=-\frac{1}{\Delta}(lg_{t\phi}-eg_{\phi\phi})}
\end{equation}
\begin{equation}\label{eqncon2}
{\displaystyle u^\phi=-\frac{1}{\Delta}(lg_{tt}+eg_{\phi t})}
\end{equation}

One can written out more explicitly as:
\begin{equation}\label{eqncon3}
{\displaystyle \frac{dt}{d\tau}=\frac{1}{\Delta}
 \left[ \left( r^2+a^2+\frac{2Ma^2}{r} \right)e-\frac{2Ma}{r}l  \right]
}
\end{equation}
\begin{equation}\label{eqncon4}
{\displaystyle \frac{d\phi}{d\tau}=\frac{1}{\Delta}
 \left[ \left( 1-\frac{2M}{r} \right)l+\frac{2Ma}{r}e  \right]
}
\end{equation}

Now applying normalization condition $g_{\mu\nu}u^\mu u^\nu=-1$ one can find the radial equation 
for equatorial motion:
\begin{equation}\label{veff1}
{\displaystyle  \frac{e^2-1}{2}=\left(\frac{dr}{d\tau}\right)^2 +V_{eff}(r) }
\end{equation}
where $ V_{eff}(r)$ is the effective potential given by:
\begin{equation}\label{veff2}
{\displaystyle V_{eff}(r)=\frac{l^2-a^2(e^2-1)}{2r^2}-\frac{M(l-ae)^2}{r^3}-\frac{M}{r} }
\end{equation}

\begin{figure}[ht]
\centerline{\includegraphics[width=6.6cm]{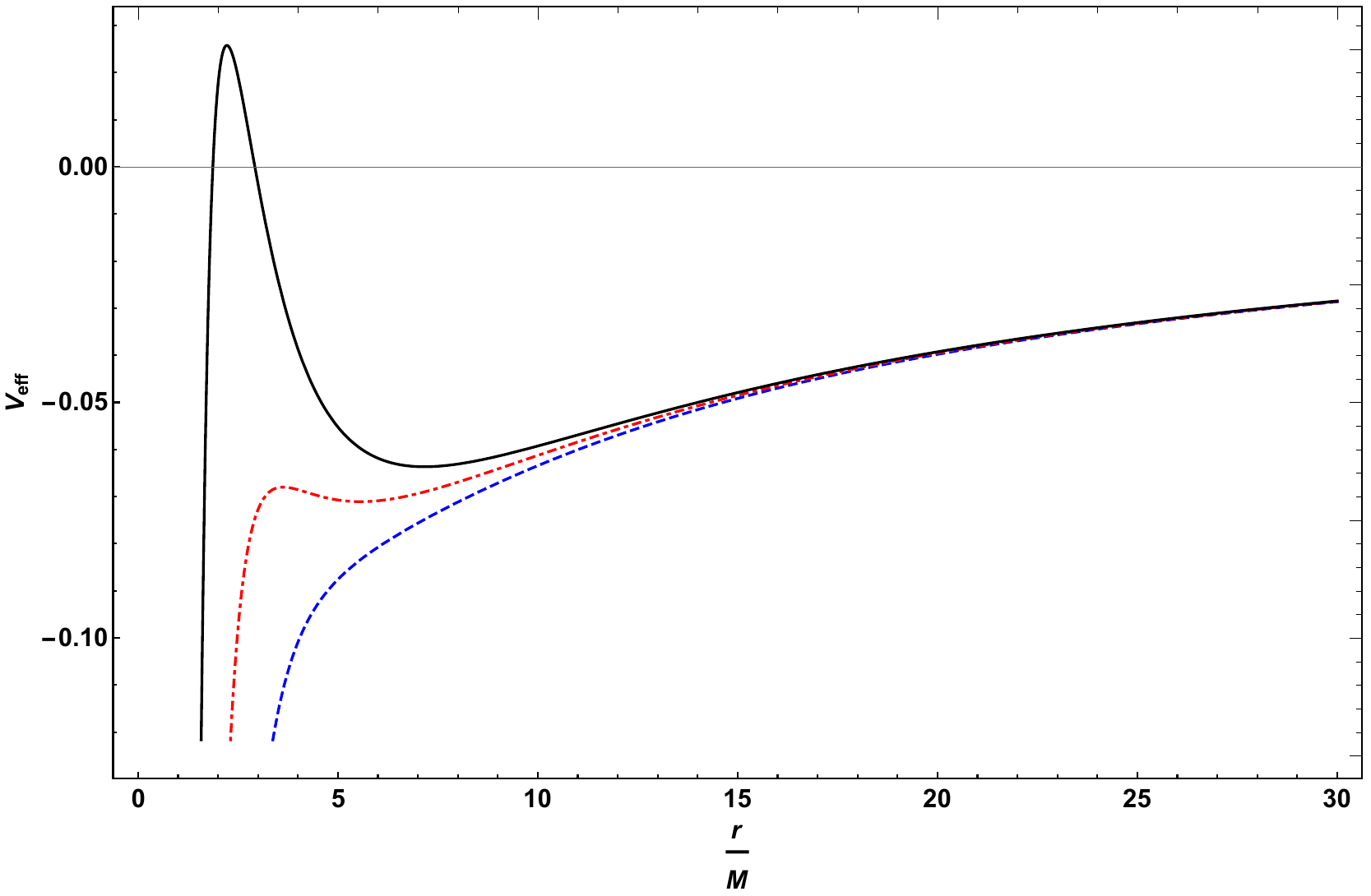} \includegraphics[width=6.6cm]{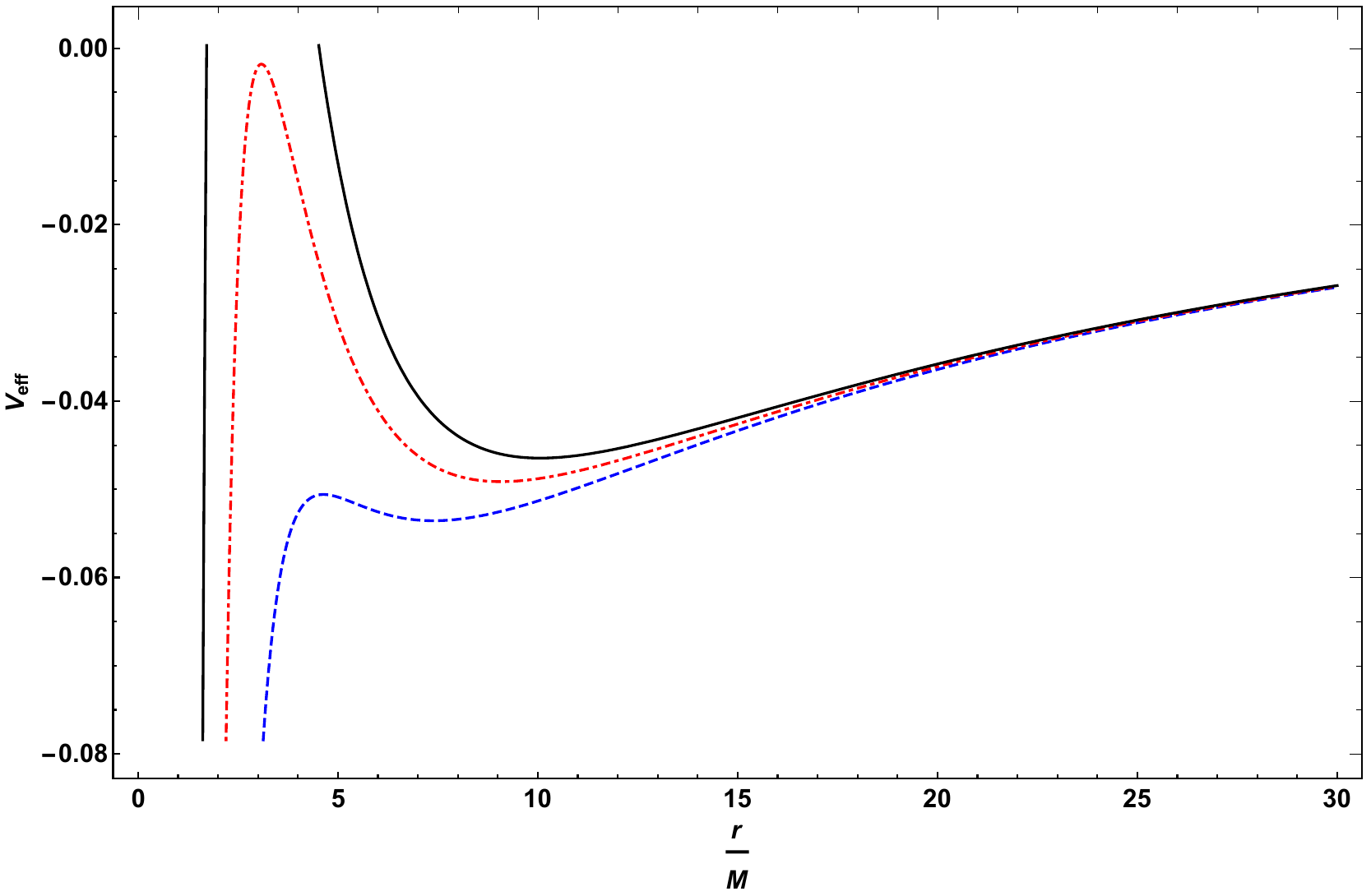}}
\vspace*{8pt}
	\caption{In the left-panel $V_{eff}$ as a function of $\frac{r}{M}$  
	 for Kerr geometry. The blue, red and black are corresponds to 
	 the value of $a=0.1M, 0.5M, 0.9M$ respectively for $\ell=3M$. 
	 In the right panel is the same for $\ell=2\sqrt 3 M$. 
	 }
\label{figrveffkerr}
\end{figure}

We've plotted the $V_{eff}$ as a function of the radial distances (in lightyear) in Fig.\ref{figrveffkerr} 
for different values of a. We see that $\frac{r}{M} \geq 15$ the values of the effective potential 
is insensitive to the rotation parameter.

\begin{figure}[ht]
\centerline{\includegraphics[width=7.5cm]{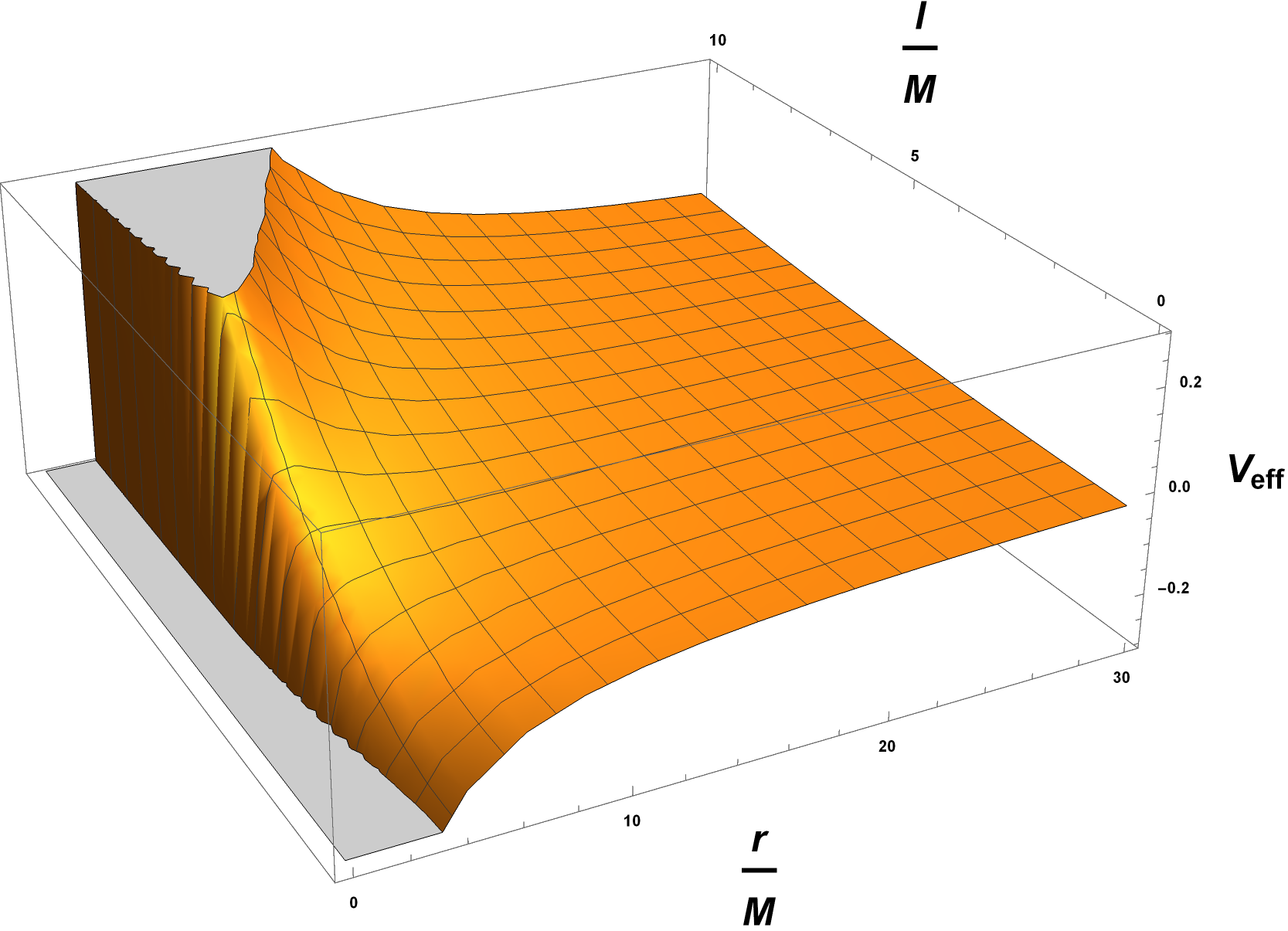}}
\vspace*{8pt}
	\caption{The $V_{eff}$ as a function of $\frac{r}{M}$ and  $\frac{\ell}{M}$ 
	(where $r$ is stands for $r_{ISCO}$ (in light years), M is in Solar mass) 
	and $\ell$ is angular momentum per unit mass.
	}
\label{figea}
\end{figure}

The three dimensional profile of effective potential is shown in Fig.\ref{figea} as a 
function of $\frac{r}{M}$ and  $\frac{\ell}{M}$. The $r$ is stands for $r_{ISCO}$ 
(in light years) and M is in Solar mass and $\ell$ is angular momentum per unit mass.

\begin{figure}[ht]
\centerline{\includegraphics[width=4.9cm]{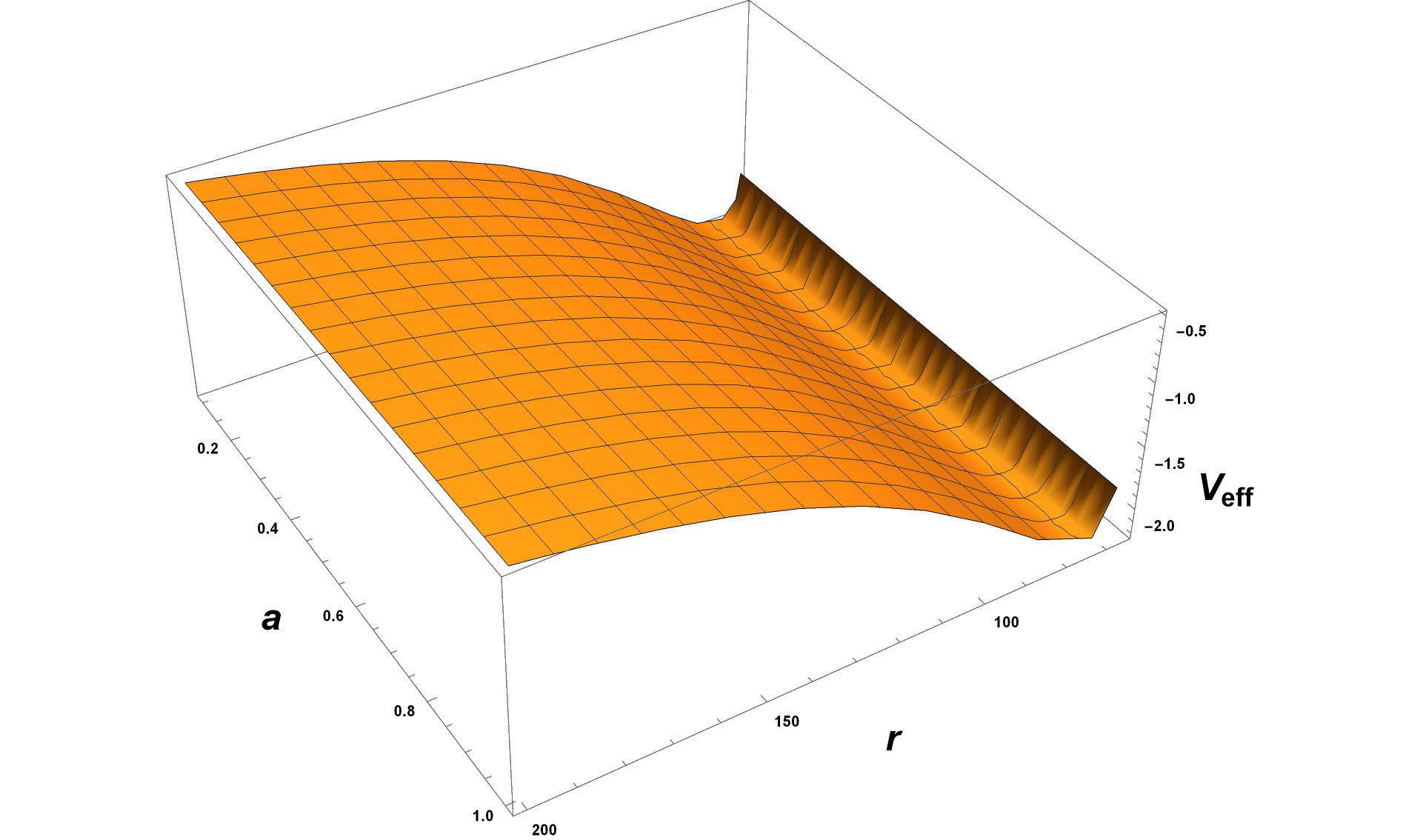} \includegraphics[width=4.9cm]{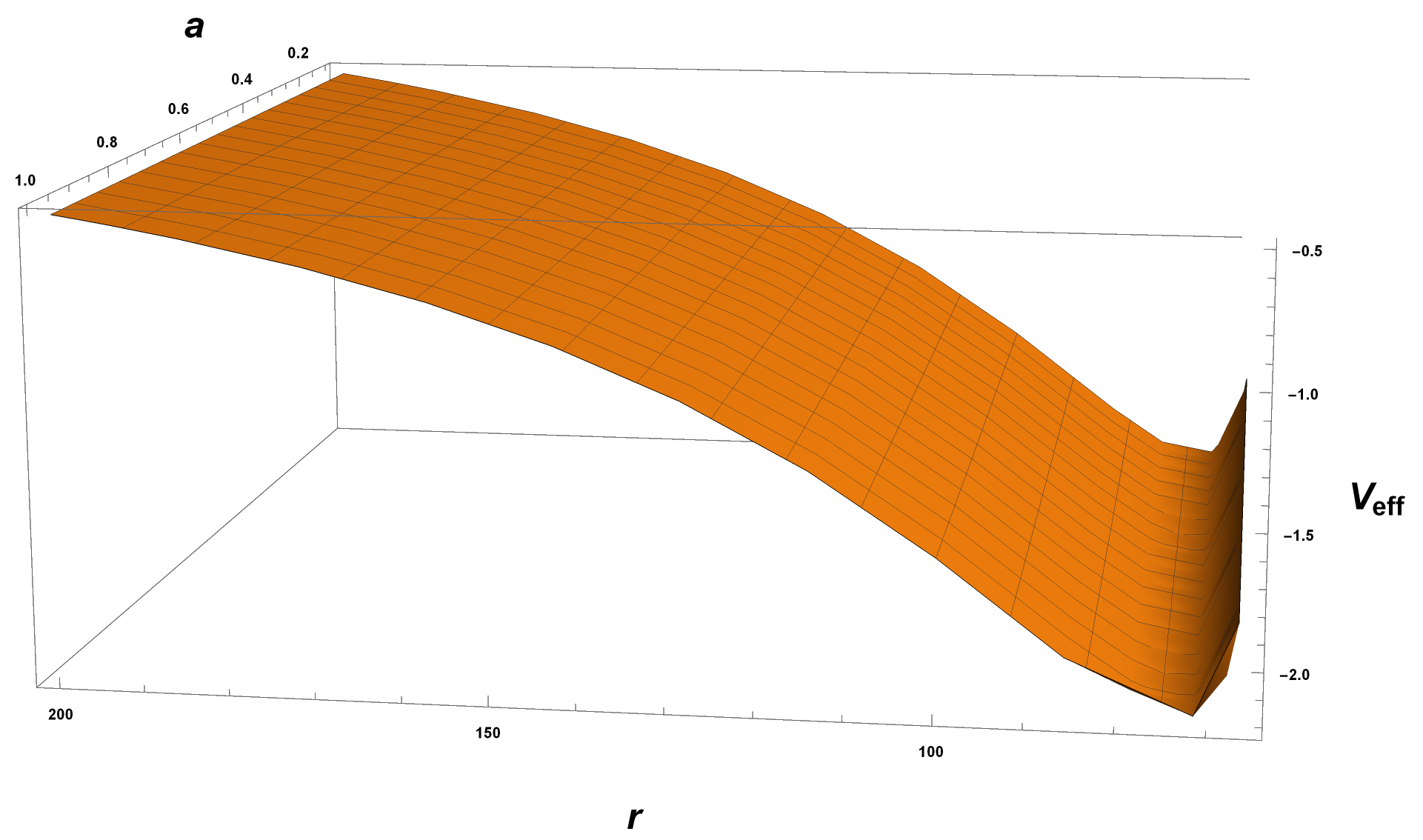} \includegraphics[width=4.9cm]{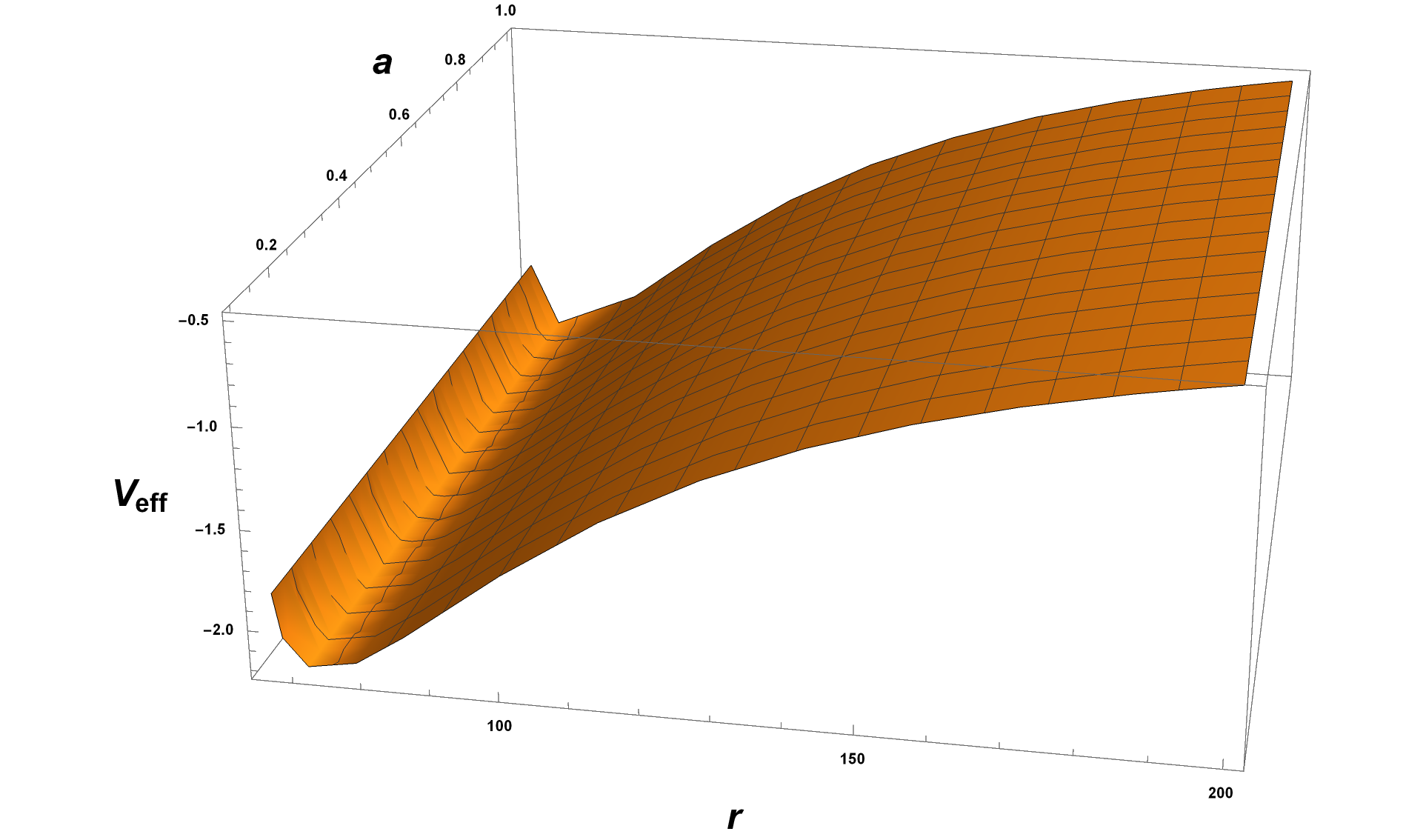} }
\vspace*{8pt}
	\caption{$V_{eff}$ as a function of $r$ and $a=\frac{\ell}{M}$ for M=100 Solar mass with 
	different orientations.}
\label{figveffarM100}
\end{figure}

In Fig.\ref{figveffarM100} we have shown three dimensional effective potential profile 
as a function of $r$ and $a=\frac{\ell}{100}$. We showed in different orientations so that 
the $v_{eff}$ profile is more clear from different angle.

\section{Solutions of Radial Equation of the Kerr Metric}

In principle we may imagine that values for $r_{ISCO}$ (radius of innermost stable circular orbit) 
are known and that the angular momentum per mass squared $\tilde{a} =\frac{a}{M}$ is an unknown 
which one wish to solve for.

Disk-accretion affects $\tilde{a}$. As black holes accrete matter and in order to do so 
expressions for the angular momentum per unit rest mass of particles ($l$) and the 
energy per unit rest mass of particles ($e$) are very important. So, solving Eqn.\ref{veff1} 
for $e$ and $\tilde{a}$ in terms of $r_{ISCO}$ one can define the potential $V(r)$ as:
$$ V(r)=-2r^2\left(V_{eff}(r) - \frac{e^2-1}{2}\right)$$

\begin{equation}\label{vr}
{\displaystyle V(r)=(r^2+a^2)(e^2-1)+\frac{2M(l-ae)^2}{r}+2Mr-l^2 }
\end{equation}

where $r>0$ and here $r_{ISCO}$ is the value of r for which we get the innermost stable, 
or marginally stable orbit with constant $r$. For $r_{ISCO}$ it necessarily follows 
$\frac{dr}{d\tau}=0$ and for particles to remain in circular orbit the conditions 
$V(r)=0$ and $V_0(r)=0$ must hold. For a stable orbit we have the condition 
$V''(r)>0$, and since $r_{ISCO}$ is the orbit which is on the verge of being unstable 
one can find that this condition becomes an equality: $V''(r)=0$. 
The derivative of $V(r)$ is given by:

\begin{equation}\label{eqnvprime}
{\displaystyle V'(r)=2r(e^2-1)-\frac{2M(l-ae)^2}{r^2}+2M }
\end{equation}

If we take the double derivative of Eqn.\ref{eqnvprime} we get 
\begin{equation}\label{eqnvdoubleprime}
{\displaystyle V''(r)=2(e^2-1)+\frac{4M(l-ae)^2}{r^3}}
\end{equation}

Now multiply of Eqn.\ref{eqnvprime} with r and add with Eqn.\ref{vr} we can get much simpler expression 
\footnote{Using computer algebraic methods one can solve energy ($e$) and angular momentum ($l$) 
in terms of $r$, $M$ and $a$(angular momentum per unit mass) \cite{press}.}

\begin{equation}\label{vrvpr}
{\displaystyle V(r)+rV'(r)=(3r^2+a^2)(e^2-1)+4Mr-l^2 }
\end{equation}

where $r>0$ and here $r_{ISCO}$ is the value of r of the innermost stable circular orbit(ISCO). As 
$r_{ISCO}$ is the minimum stable point so $V'(r)=0$ and for the innermost stable circular orbit 
in the equatorial plane the second derivative $V''(r)=0$, therefore we can differentiate 
Eq.\ref{vrvpr} second time and get 

\begin{equation}\label{drvr}
{\displaystyle \frac{d}{dr}[V(r)+rV'(r)]=2V'(r)+rV''(r) }
\end{equation}

The above equation only contains the first and second derivative. 

As soon as we have imposed the condition $V''(r) = 0$ the radius $r$ must 
necessarily be $r_{ISCO}$. Hereafter whenever we use $r$ it always 
referring to $r_{ISCO}$.

\begin{figure}[ht]
\centerline{\includegraphics[width=6.6cm]{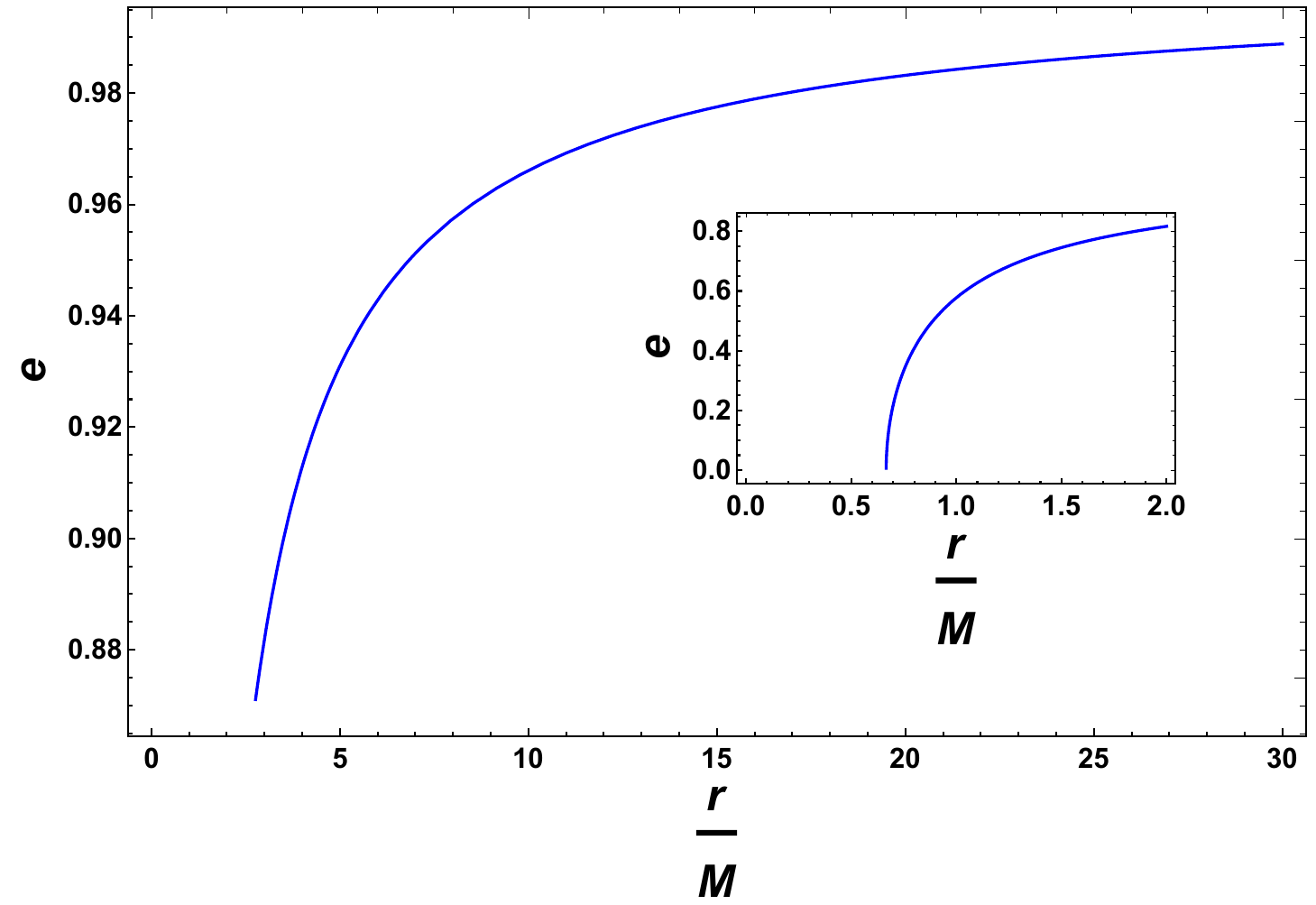} \includegraphics[width=6.6cm]{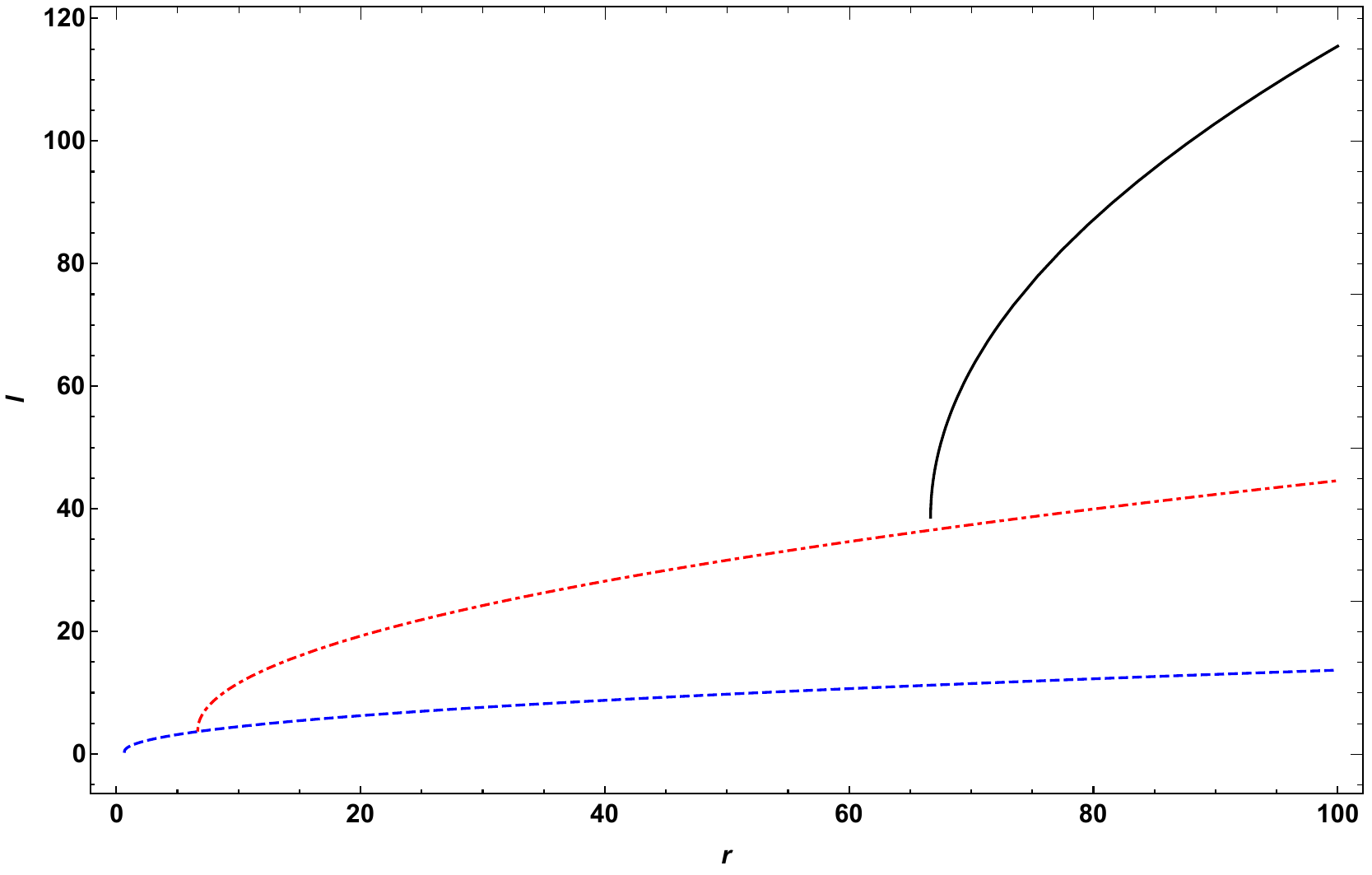} }
\vspace*{8pt}
	\caption{In the left-panel $e$ (energy per unit rest mass) is plotted as a function 
of $\frac{r}{M}$ following Eqn.\ref{emr}.  Please note that the values of $\frac{r}{M} \sim 0.66$ for real $e$. 
In the right-panel $\ell$ is plotted as a function of $r$ for M=1(dashed blue), M=10 (dotted-dash red) 
and M=100(solid black). The finite value of $\ell$ is starting from 0.66, 6.6 and 66.6 for M=1,10,100 respectively.}
\label{figerm}
\end{figure}

With this in mind we find that Eq.\ref{drvr} reduces to:

\begin{equation}\label{drvrb}
{\displaystyle \frac{d}{dr}[V(r)+rV'(r)]=6r(e^2-1)+4M=0}
\end{equation}

and from this it follows that

\begin{equation}\label{emr}
{\displaystyle e=\left(1-\frac{2M}{3r}\right)^{\frac{1}{2}} }
\end{equation}

considering only the positive solution. The energy ($e$) per unit rest mass as a function of $\frac{r}{M}$ 
is plotted in the left-panel of Fig.\ref{figerm}.

In the right-panel of Fig.\ref{figerm} we've plotted $\ell$ as a function of $r$ for M=1,10,100 Solar mass. 
It is clear from the figure that the finite values of $\ell$ is started from 0.66, 6.6 and 66.6 for M=1,10,100 
respectively shows that it simply scales.

For the critical point Eqn.\ref{eqnvdoubleprime} is equal to zero and from Eqn.\ref{emr} 
one can get the expression for $\ell$

\begin{equation}\label{ellarm}
{\displaystyle l=a\sqrt(1-\frac{2M}{3r}) \pm {\frac{r}{\sqrt3}}}
\end{equation}

From Eqn.\ref{vrvpr} and Eqn.\ref{emr} and doing simple algebra one can get 

\begin{equation}\label{ellsq}
{\displaystyle l^2=2M(r-\frac{a^2}{3r})}
\end{equation}

Now from Eqn.\ref{ellarm} and Eqn.\ref{ellsq} with simple algebra one can get 
second order equations of $a$: 

\begin{equation}\label{asq}
	{\displaystyle a^2 \pm {\frac{2r}{\sqrt 3}} \sqrt (1- {\frac{2M}{3r}}) a - rM + \frac{r^2}{6}=0}
\end{equation}

The solution of this Eqn.\ref{asq} is

\begin{equation}\label{arm}
{\displaystyle a=\pm \frac{4\sqrt{Mr}}{3}\mp \frac{r}{\sqrt{3}}\sqrt{\left(1-\frac{2M}{3r}\right)}}
\end{equation}

By defining $\tilde{a}=\frac{a}{M}$ Eq.\ref{arm} can be simplified as:
\begin{equation}\label{atilde}
	{\displaystyle \tilde{a}=\pm \frac{4}{3}\sqrt{\frac{r}{M}} \mp \frac{r}{M\sqrt 3} \sqrt{1- \frac{2M}{3r}}}
\end{equation}

where the upper signs refer to corotation and the lower signs refer to counter rotation.

\begin{figure}[ht]
\centerline{\includegraphics[width=7.1cm]{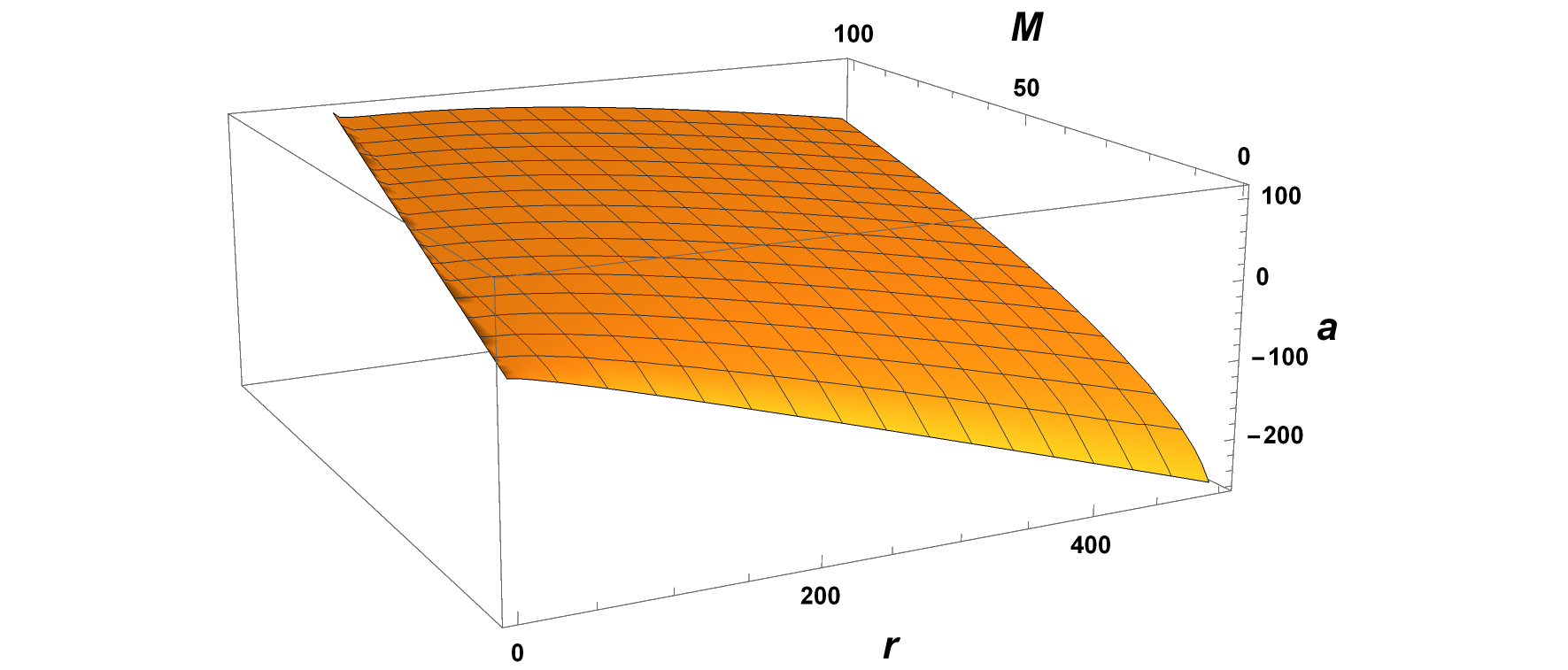} \includegraphics[width=7.1cm]{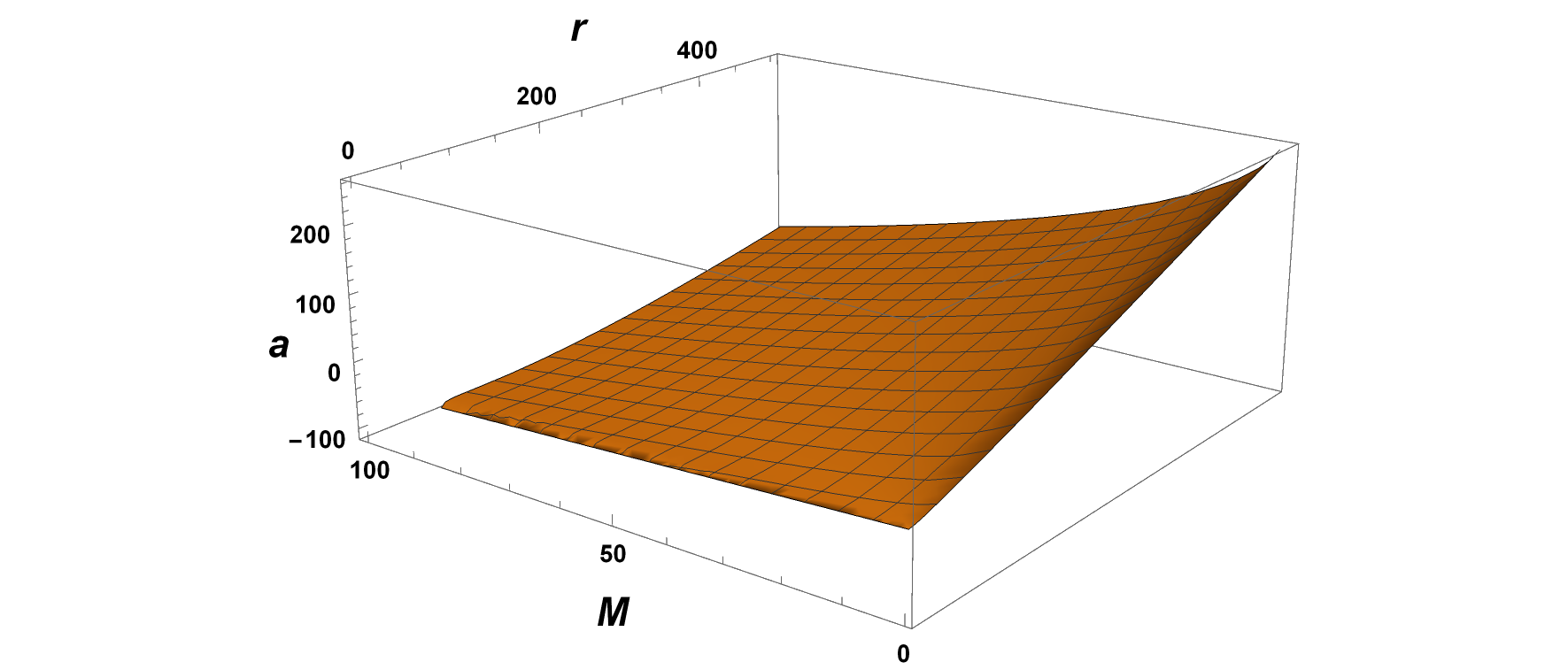}}
\vspace*{8pt}
	\caption{The three dimensional profile of $a$ as a function of $r=[0:10]$ and $M=[1-100]$ (in unit 
	of Solar mass) with positive (left-panel) and negative(right-panel) sign for the first-term of Eqn. .
         }
\label{fig_arM}
\end{figure}

We've shown in Fig.\ref{fig_arM} the three dimensional profile of $a$ as a function of $r$ and $M$ with 
positive (left-panel) and negative (right-panel) sign for the first-term of Eqn.24.

The profile of $\tilde{a}$ as a function of  $\frac{r}{M}$ is plotted in the 
right-panel of Fig.\ref{figea}.

\begin{figure}[ht]
\centerline{\includegraphics[width=6.6cm]{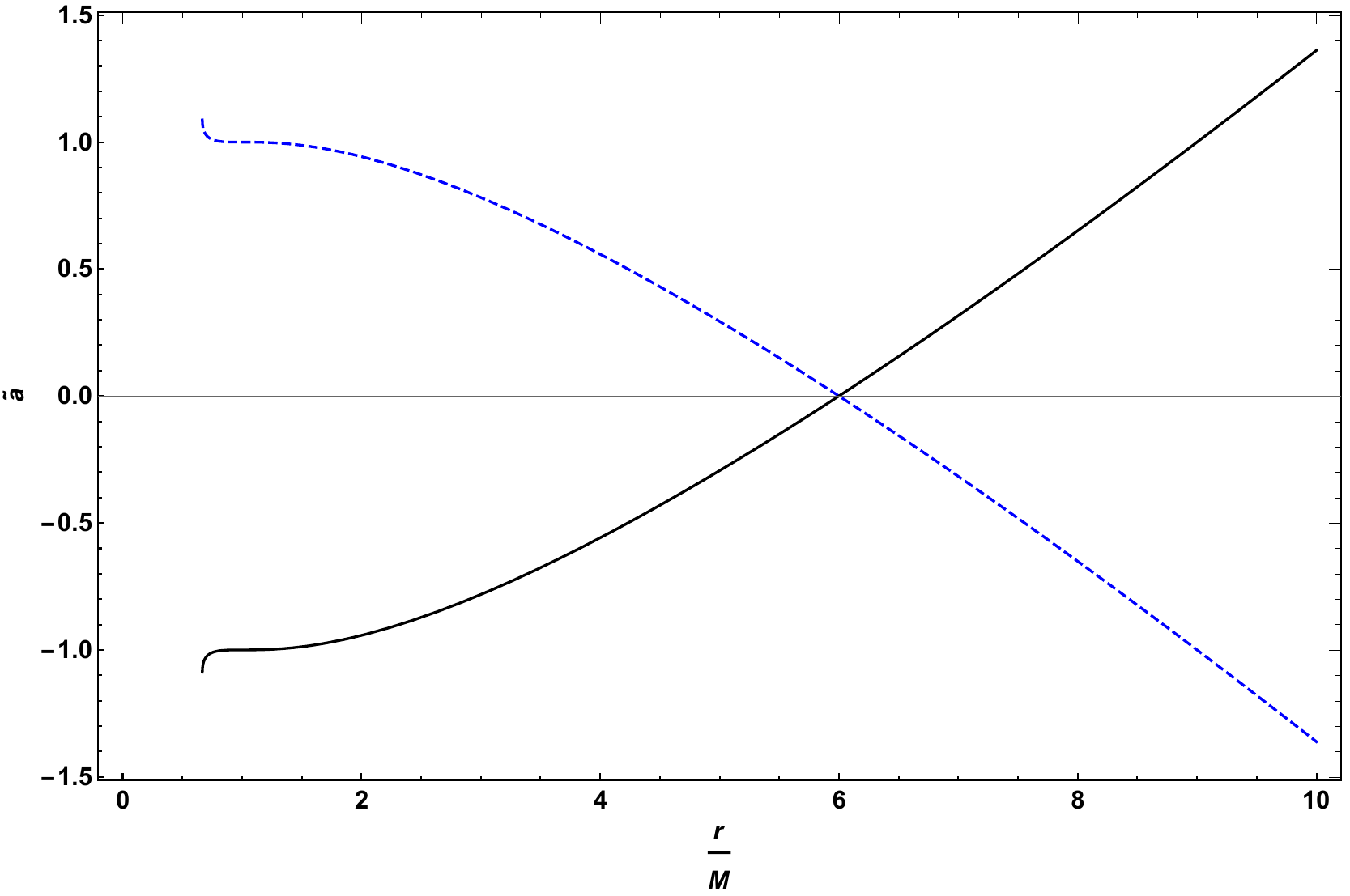} \includegraphics[width=6.6cm]{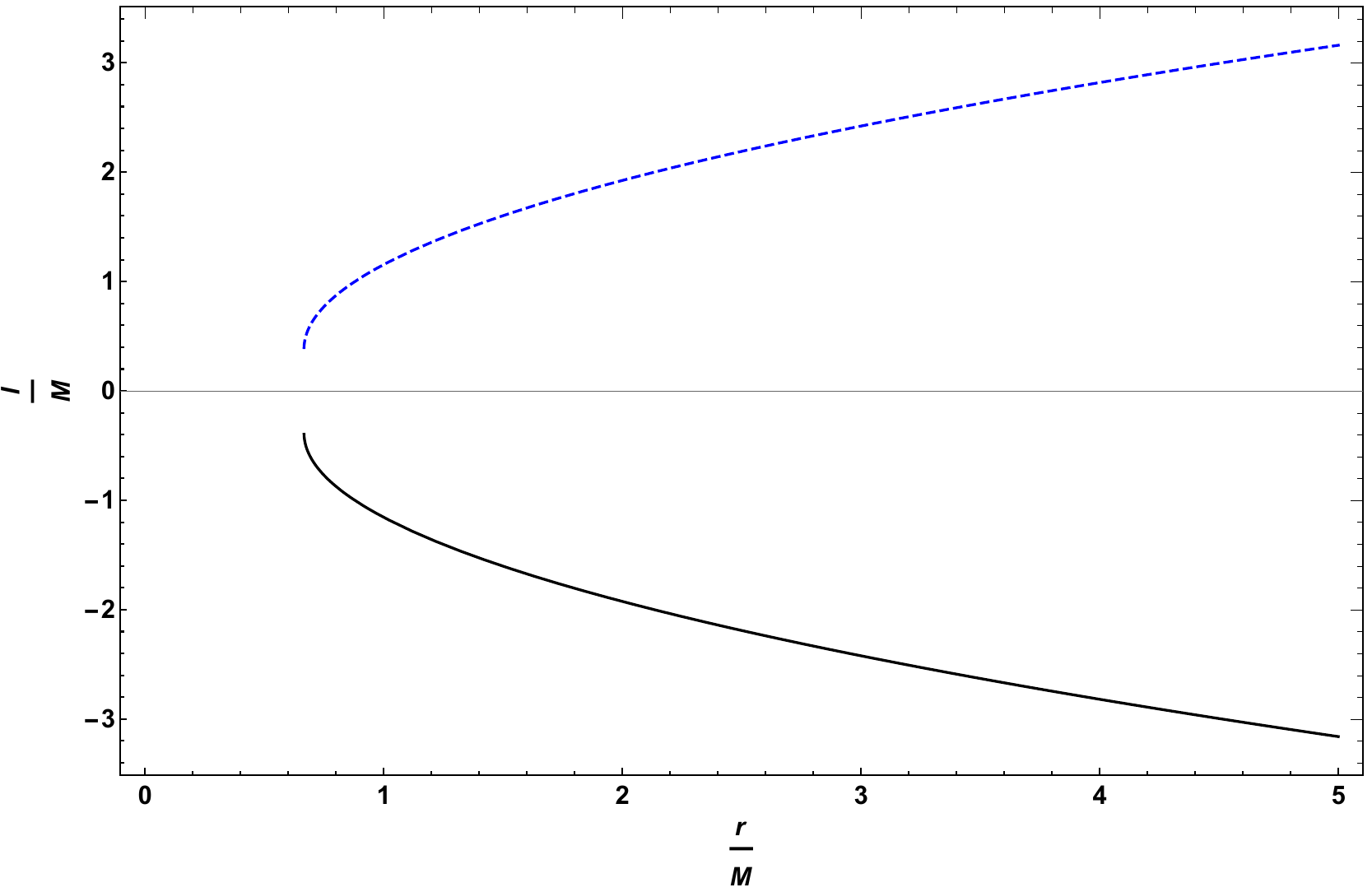} }
\vspace*{8pt}
	\caption{$\tilde a$ as a function of $\frac{r}{M}$ following Eqn.25 in the left-panel. 
	 In the right-panel $\frac{\ell}{M}=\frac{J}{M^2}$ is plotted as a function of $\frac{r}{M}$ 
	 following Eqn.26 taking +ve sign only.
	 }
\label{fig_atildex_jm2x}
\end{figure}

We've plotted the angular momentum per unit mass square mass, i.e., $\tilde a = \frac{\ell}{M^2} $ as 
a function of $\frac{r}{M}$ following Eqn.25 in the left-panel of Fig.\ref{fig_atildex_jm2x}. It is clear that 
a counter rotates particles starts to corotate with Black hole once $\frac{r}{M} \geq 6$ and vice-versa. 
We can say that is the outer most boundary of ergosphere for the upper sign of $\tilde a$. 

In the right-panel of Fig.\ref{fig_atildex_jm2x} we plotted Eqn.26 (i.e., the angular momentum per unit mass) 
as a function of $\frac{r}{M}$. Lower than the values of $\frac{r}{M} \leq 0.66$ is not physically not possible 
because $M$ is increased very much so to rotate the black hole needs huge energy say nearly infinity and is 
not possible physically.

\begin{figure}[ht]
\centerline{\includegraphics[width=4.9cm]{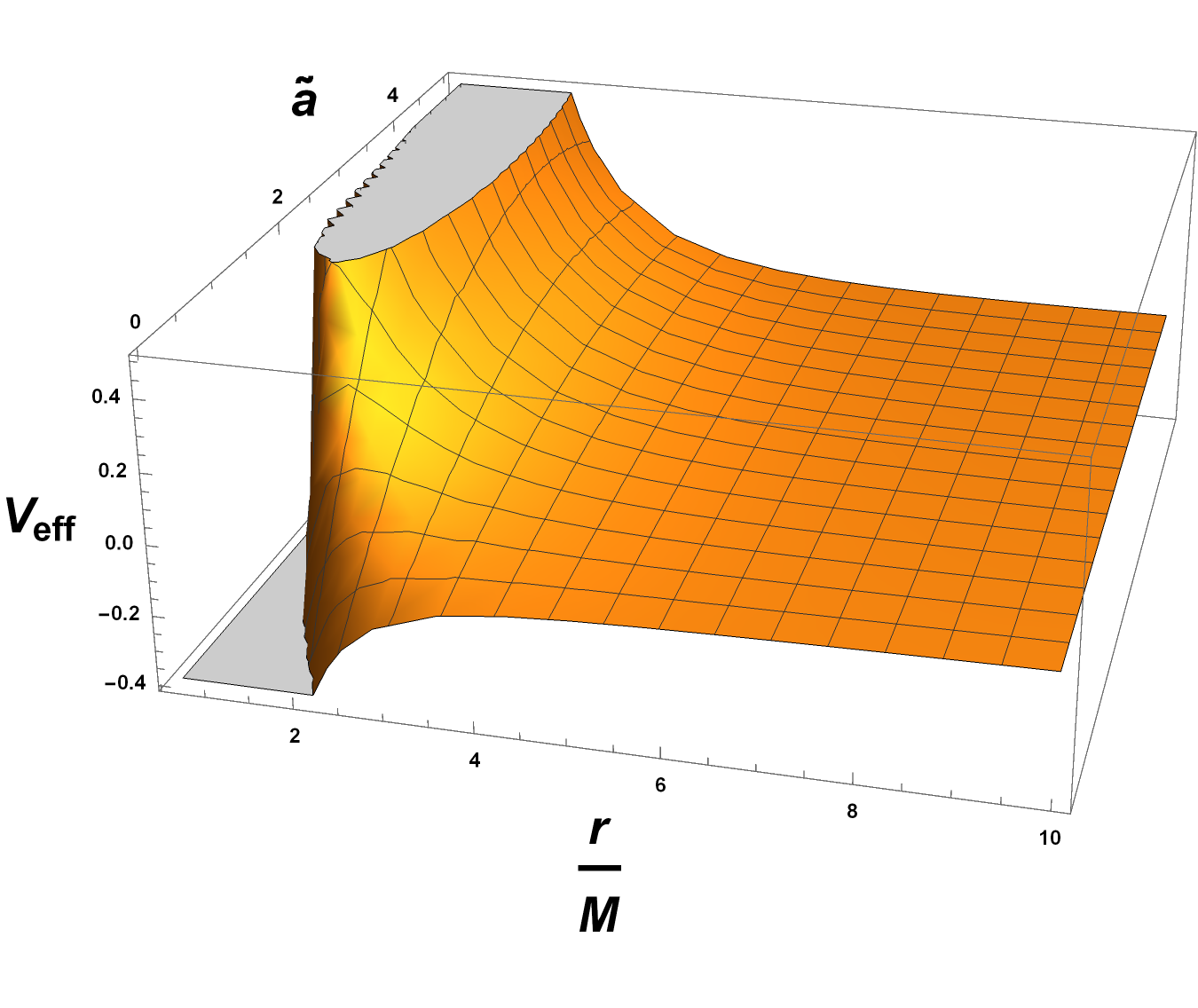} \includegraphics[width=4.9cm]{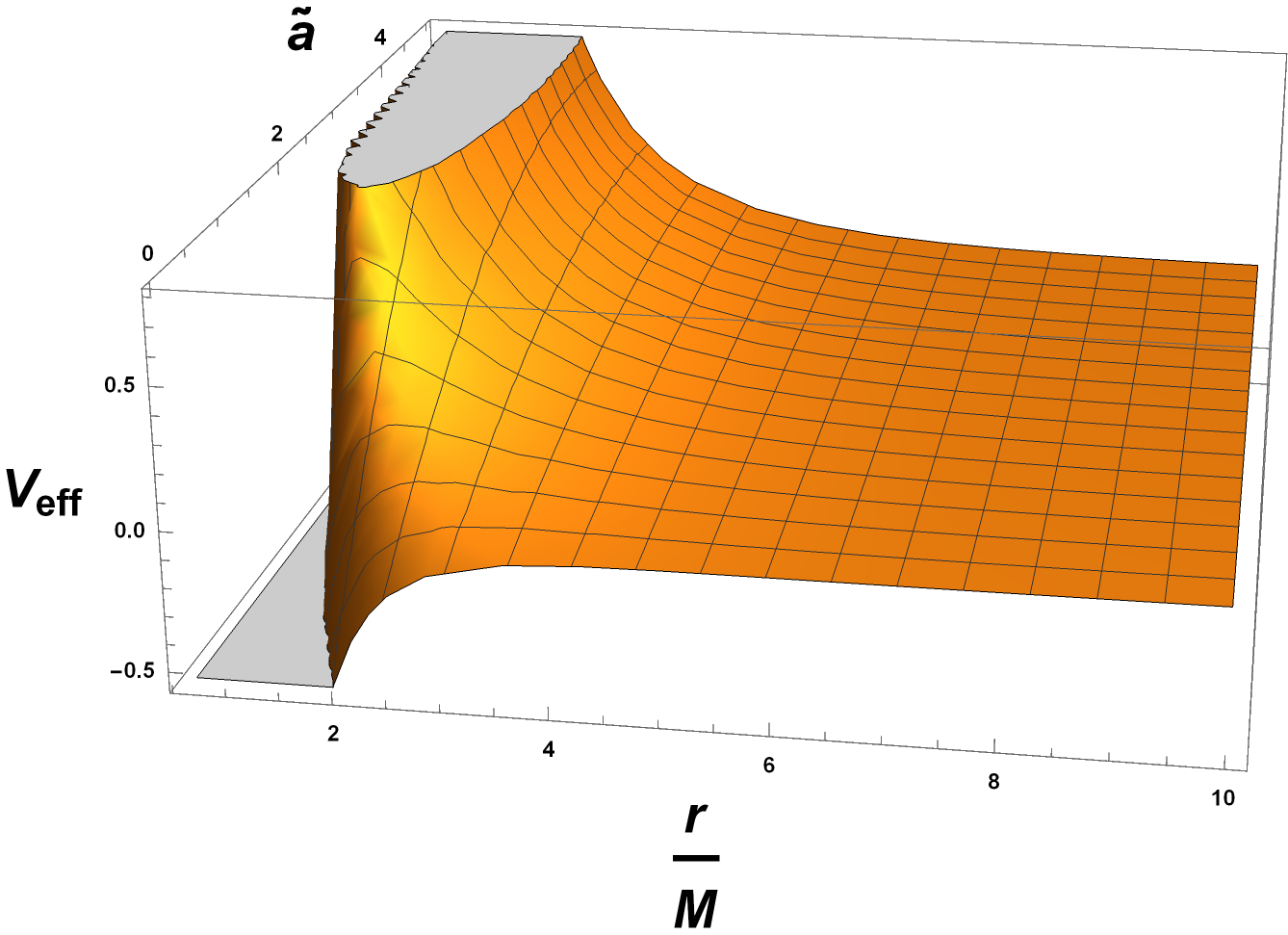} \includegraphics[width=4.9cm]{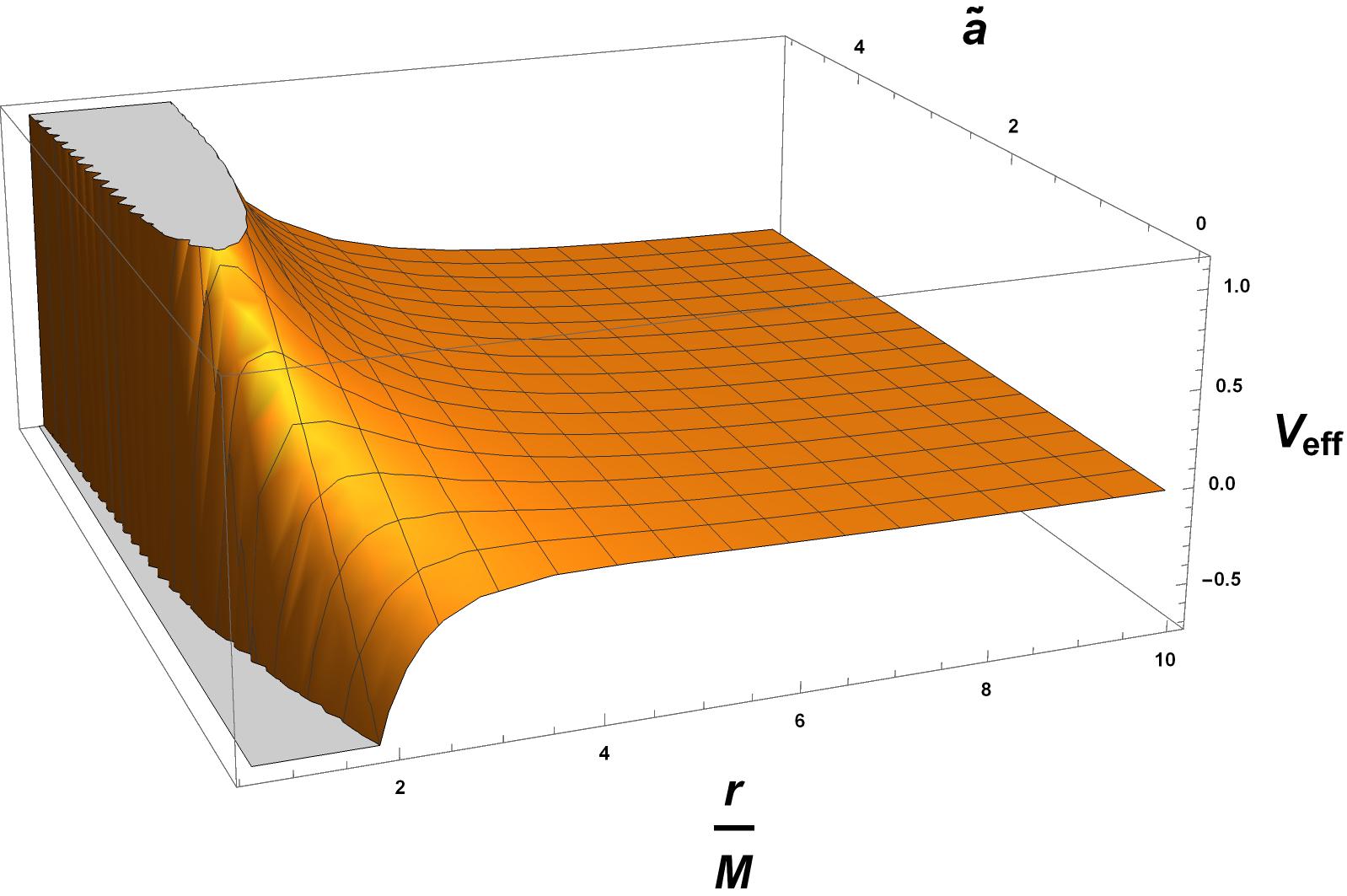} }
\vspace*{8pt}
	\caption{$V_{eff}$ as a function of $\frac{r}{M}$ and $\tilde a = \frac{a}{M}$   
	 for $\frac{\ell}{M}$= 3, 2 $\sqrt 3$ and 4 respectively for left, middle and right 
	 panel. Please note that the values $V_{eff}$ is zero if $\frac{r}{M} \leq 0.66$ for 
	 all values of $\tilde a = \frac{a}{M}$.}
\label{figveff_rm_am}
\end{figure}

In Fig.\ref{figveff_rm_am} we've plotted $V_{eff}$ as a function of $\frac{r}{M}$ and 
$\tilde a = \frac{a}{M}$ for $\frac{\ell}{M}$= 3, 2 $\sqrt 3$ and 4 respectively for left, middle and right panel. 
It is clear that the values $V_{eff}$ is zero if $\frac{r}{M} \leq 0.66$ for all values of $\tilde a = \frac{a}{M}$.
We see that for a particular value of $\tilde a $ the value of $V_{eff}$ create the barrier in small region 
of $\frac{r}{M}$ which prevent the object to collapse into black hole. The second term of $V_{eff}$ is 
dominant with respect to the first term for the low values of $\frac{r}{M}$.

\section{Numerical Analysis}

Finally we expresses $\ell$ as a function of $r$ using the expressions of $\tilde{a}$ and $e$. 
Using Eq.\ref{ellarm} we solved for $\ell$. Taking $a=\tilde{a}M$ and the upper sign and with 
simplification leads to 

\begin{equation}\label{ellarmz}
{\displaystyle l=\frac{2M}{3^\frac{3}{2}}\left[1+2\left(\frac{3r}{M}-2\right)^\frac{1}{2}\right]}
\end{equation}

We have now derived three expressions for $e$, $l$ and $\tilde{a}$ in terms of $z$, where $z=\frac{r_{ISCO}}{M}$ 
is called the Boyer-Lindquist radius~\cite{boyer}.

\begin{equation}\label{ez}
{\displaystyle e=\left(1-\frac{2}{3z}\right)^{\frac{1}{2}} }
\end{equation}

\begin{equation}\label{lzm}
{\displaystyle l(z.M)=\frac{2M}{3^\frac{3}{2}}\left[1+2\left(3z-2\right)^\frac{1}{2}\right]}
\end{equation}

\begin{equation}\label{atildez}
{\displaystyle \tilde{a}=\pm \frac{1}{3}\sqrt{z}\left[4- \sqrt{\left(3z-2\right)}\right]}
\end{equation}

We mention here that $z$ can be solved for as a function of $\tilde{a}$ following \cite{press}:

\begin{equation}\label{zatilde}
{\displaystyle z(\tilde{a})=3+w_2 \mp \left[(3-w_1)(3+w_1+2w_2)\right]^\frac{1}{2}}
\end{equation}

where 

\begin{equation}\label{wone}
{\displaystyle w_1=1+(1-\tilde{a}^2)^\frac{1}{3}\left[(1+\tilde{a})^\frac{1}{3}+(1-\tilde{a})^\frac{1}{3}\right]}
\end{equation}

\begin{equation}\label{wtwo}
{\displaystyle  w_2=(3\tilde{a}+w_{1}^{2})^\frac{1}{3}}
\end{equation}


The upper sign in Eqn.\ref{zatilde} refers to corotation and the lower sign refers to counter rotation.

\begin{figure}[ht]
\centerline{\includegraphics[width=6.6cm]{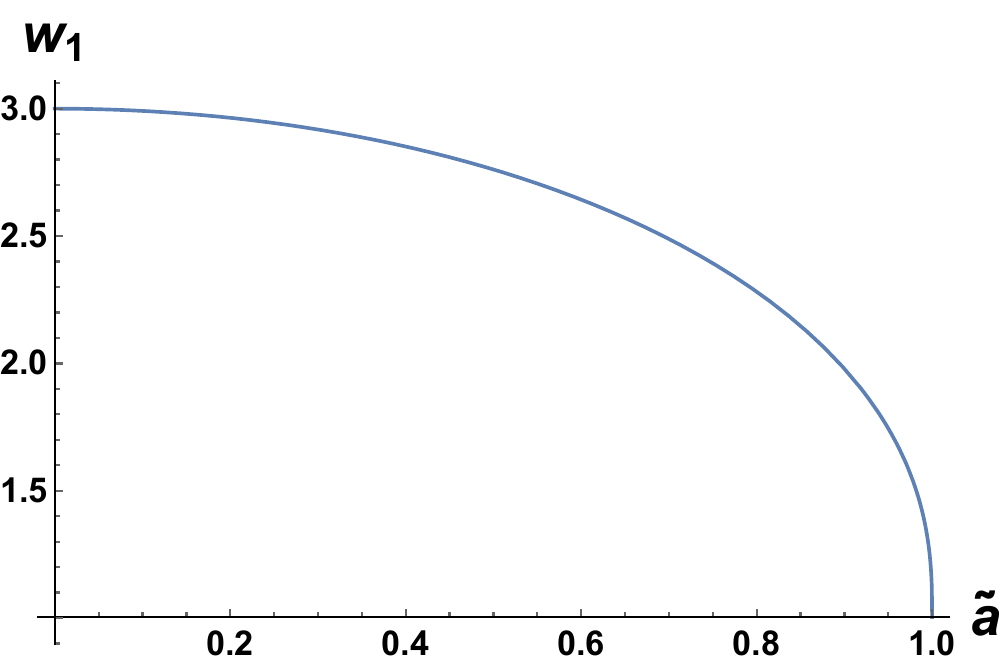} \includegraphics[width=6.6cm]{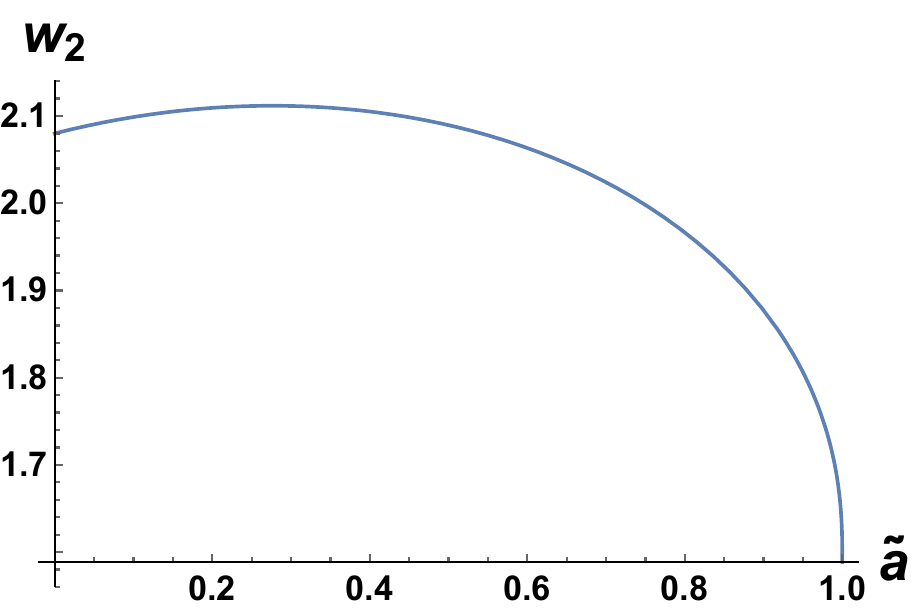}}
\vspace*{8pt}
	\caption{$\tilde a$ as a function of $w_1$ following Eqn. in the left-panel. 
	 In the right-panel $\tilde a$  as a function of $w_2$ following Eqn.
	 }
\label{fig_atildew1w2}
\end{figure}

\begin{figure}[ht]
\centerline{\includegraphics[width=6.6cm]{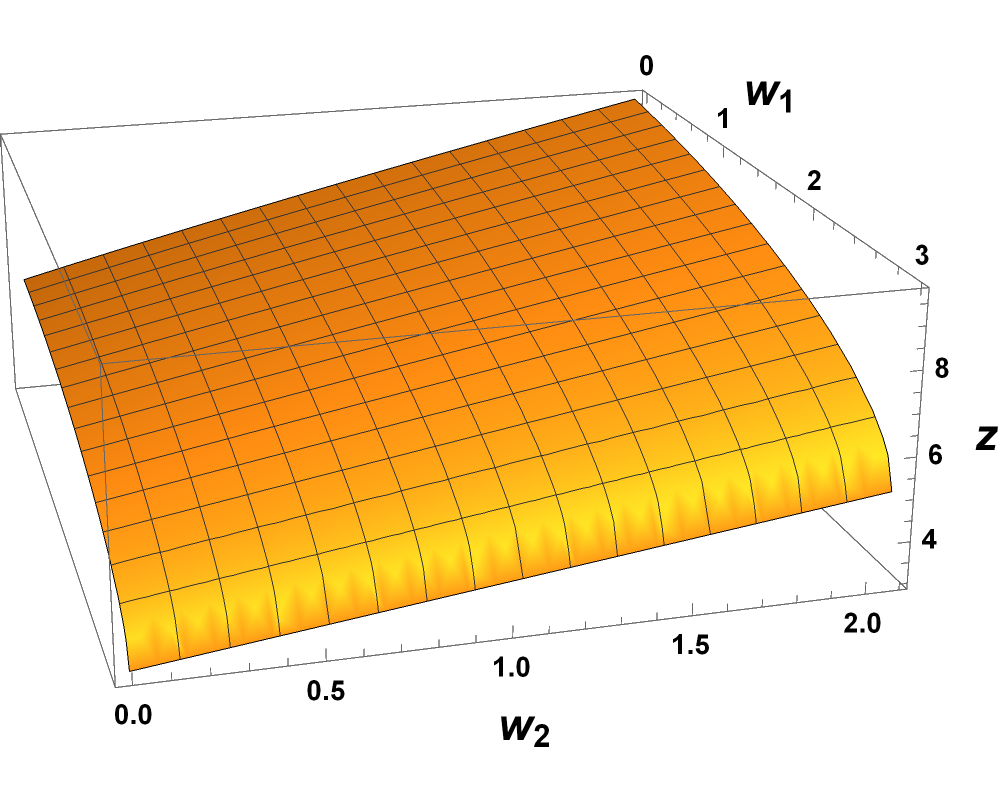} \includegraphics[width=6.6cm]{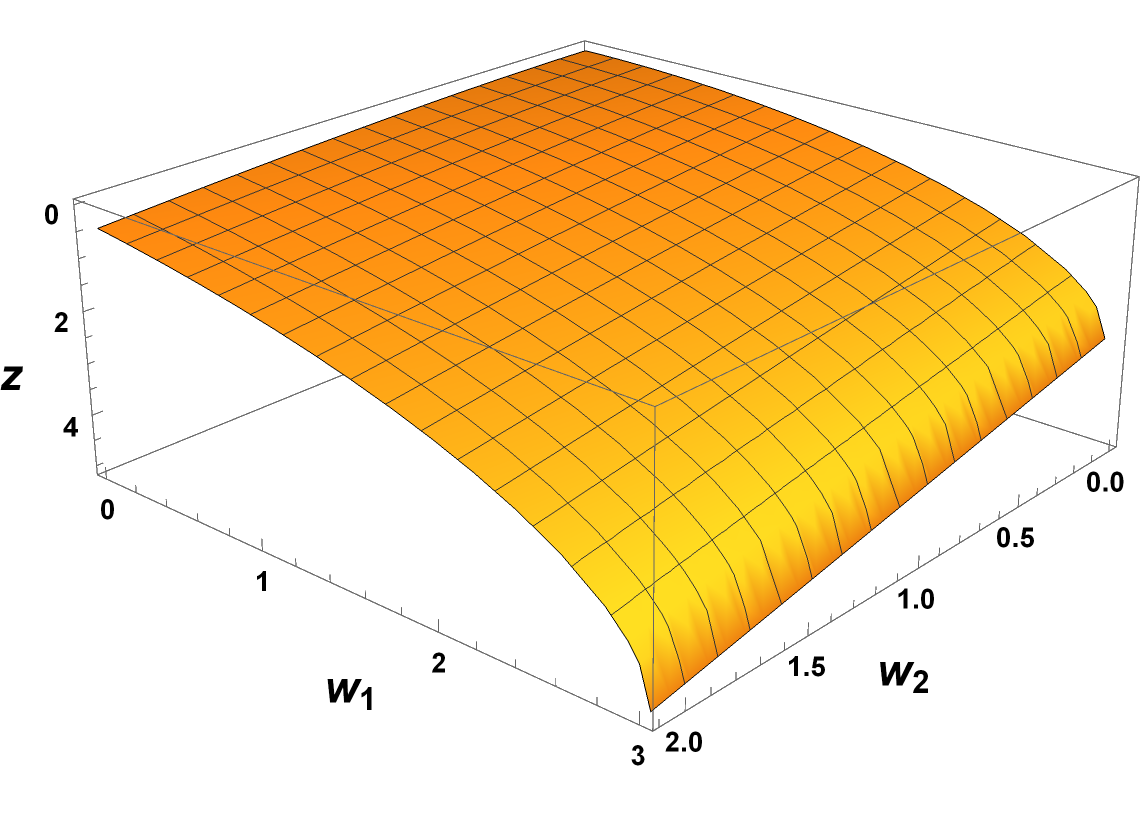}}
\vspace*{8pt}
	\caption{The three dimensional profile of $z(\tilde a)$ as a function of 
	$w_1$ and $w_2$ with positive (left-panel) 
	and negative(right-panel) sign convention following Eqn... The maximal values of 
	$w_1$ and $w_2$ are follow from Fig.\ref{fig_atildew1w2}.
	 }
\label{fig_ztildeaw1w2}
\end{figure}

In Fig.\ref{fig_atildew1w2} we've shown the variations of $\tilde a$ as a function of $w_1$ and $w_2$ 
in the left-panel and right-panel respectively. This shows the maximal allowed values of $w_1$ and $w_2$ 
in the Schwarzschild limit. This limiting values of  $w_1$ and $w_2$ have been taken into 
considerations while plotting the three dimensional profile of $z(\tilde a)$, i.e., the Boyer-Lindquist radius
in Fig.\ref{fig_ztildeaw1w2}. 

\begin{figure}[ht]
\centerline{\includegraphics[width=6.6cm]{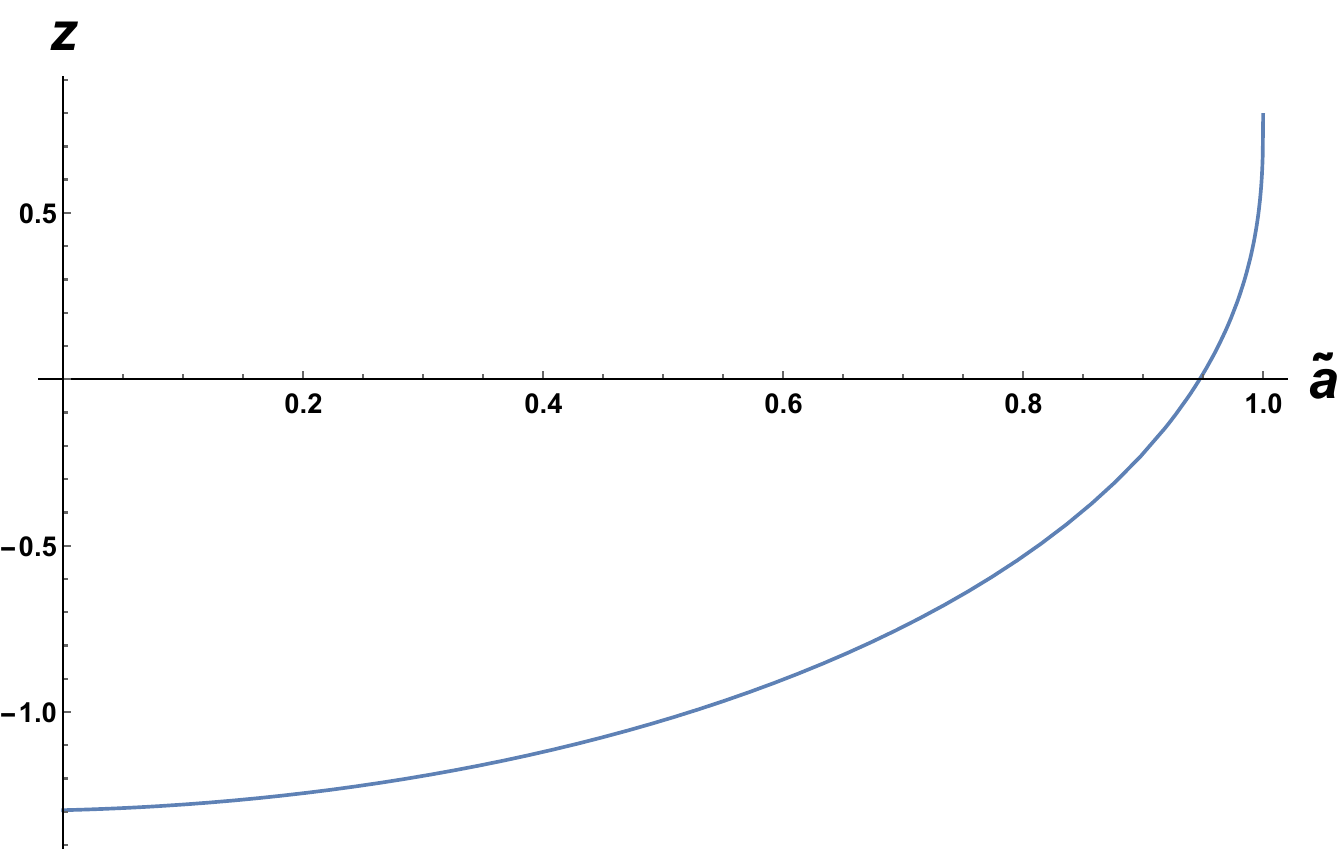} \includegraphics[width=6.6cm]{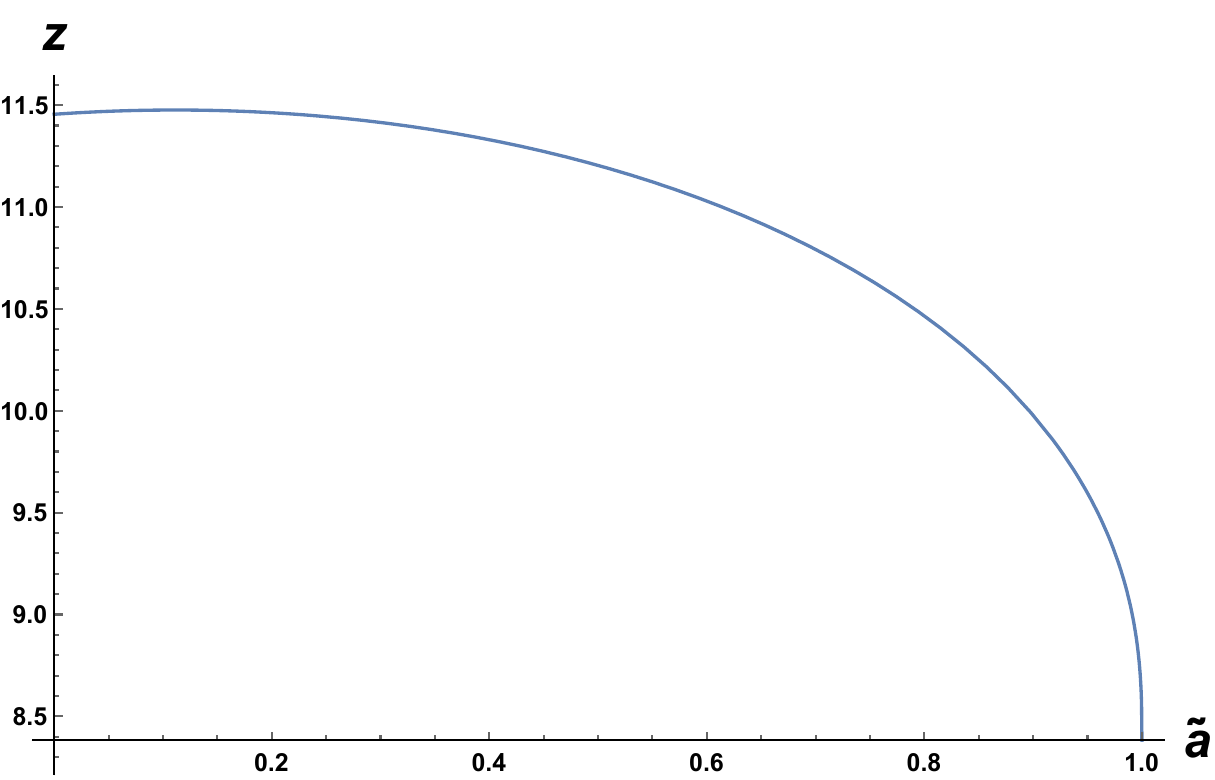}}
\vspace*{8pt}
	\caption{$z(\tilde a)$ as a function of $\tilde a$ following Eqn.29 with negative (left-panel) 
	and positive(right-panel) signature. 
	 }
\label{fig_zw1w2}
\end{figure}

We've shown the Boyer-Lindquist radius ($z$), i.e., the radius of rotating particle around the black hole, 
as a function of $\tilde a$ following Eqn.29 with negative and positive sign convention in the left 
and right-panel of Fig.\ref{fig_zw1w2} respectively considering $M=1$. The Boyer-Lindquist radius is maximum 
for $\tilde a = 0$ and this is equivalent to the Schwarzschild case. The space time fabric is directly 
converges towards black hole -- so energy required to rotate around black hole is the kinetic energy of the particle itself. 
Now once $\tilde a$ increase the space time fabric get twisted so as a result the geodesic path of 
particle also twisted around the BH. In this scenario the particle need less amount of 
kinetic energy for the stable orbit as compared to Schwarzschild case.
For $\tilde a$ =1 the maximum value of innermost stable orbit nearly $z \simeq 8.4(1.0)$ for 
positive and negative signature in Eqn.29.

In Fig.\ref{fig_ztildeaw1w2} we shown the three dimensional profile of $z(\tilde a)$ 
with $w_1$ and $w_2$ with positive (left-panel) and negative (right-panel) sign 
convention following Eqn.30. The allowed maximal values of $w_1$ and $w_2$ are taken 
from Fig.\ref{fig_atildew1w2}.

\section{Summary and Conclusion}

The BH naturally emerges from the metric solution of the general relativity. The rotating BH 
as known as Kerr and general relativity put an upper limit on the angular momentum per mass 
squared of black holes $\leq 1$. If the value is more than one then above which the event 
horizon of the Kerr BH is not exist. In this paper we have studied the effective potential 
of the rotating Kerr metric and shown the effective potential profile for different values 
of rotation parameter ($a$). The solution of the radial equation of the Kerr metric 
is found and hence of the angular momentum per unit mass squared $\tilde a$. 
The behaviors of $\tilde a$, the energy per unit rest mass ($e$) are shown as a function of 
$\frac{r}{M}$. We demarcated the maximum values of radius of innermost stable circular orbit, 
i.e., $r_{ISCO}$, for $\tilde a=1$. Our numerical findings are relevant while studying the 
BH's accretion of the matter from surrounding areas, consequently increasing the mass and the  
angular momentum of the BHs. The solutions of the angular momentum equations are 
very important to study the accreation of the matter into the BHs and we will address 
this issues in other study.

 

\begin{thebibliography}{0}    

\bibitem{Vagnozzi:2022moj}
S.~Vagnozzi, R.~Roy, Y.~D.~Tsai, L.~Visinelli, M.~Afrin, A.~Allahyari, P.~Bambhaniya, D.~Dey, S.~G.~Ghosh and P.~S.~Joshi, \textit{et al.}
Class. Quant. Grav. \textbf{40}, no.16, 165007 (2023)
doi:10.1088/1361-6382/acd97b
[arXiv:2205.07787 [gr-qc]].

\bibitem{hartle} James B. Hartle, Gravity: An Introduction to Einstein’s General Relativity,
Addison Wesley, ISBN: 0–8053–8662–9, 2003.

\bibitem{weinberg} Steven Weinberg, Gravitation and Cosmology: Principles and Applications of the Gen-
eral Theory of Relativity, John Wiley \& Sons, ISBN: 0–471–92567–5, 1972.

\bibitem{kerr} Roy Patrick Kerr, Gravitational Field of a Spinning Mass as an Example of Alge-
braically Special Metrics, Phys. Rev. Letters, Vol. 11, p. 237, 1963.

\bibitem{Donmez:2023kmh} O.~Donmez,
[arXiv:2307.11725 [astro-ph.HE]].

\bibitem{Donmez:2022dze}
O.~Donmez, F.~Dogan and T.~Sahin,
Universe \textbf{8}, no.9, 458 (2022)
doi:10.3390/universe8090458
[arXiv:2205.14382 [astro-ph.HE]].

\bibitem{Managave:2023rhn}
K.~G.~Managave, H.~A.~Redekar, R.~B.~Kumbhar, S.~P.~Das and K.~Y.~Rajpure,
Int. J. Mod. Phys. A \textbf{38}, no.28, 2350152 (2023)
doi:10.1142/S0217751X2350152X
[arXiv:2303.07736 [gr-qc]].

\bibitem{Redekar:2023qra}
H.~A.~Redekar, R.~B.~Kumbhar, S.~P.~Das and K.~Y.~Rajpure,
[arXiv:2308.12639 [gr-qc]].

\bibitem{bardeen} James M. Bardeen, Kerr Metric Black Holes, Nature, Vol. 226, p. 64, 1970.

\bibitem{press} James M. Bardeen, William H. Press and Saul A. Teukolsky, Rotating Black Holes:
Locally Nonrotating Frames, Energy Extraction, and Scalar Synchrotron Radiation,
The Astrophysical Journal, Vol. 178, p. 347, 1972.

\bibitem{carter} Brandon Carter, Global Structure of the Kerr Family of Gravitational Fields, Physical
Review, Vol. 174 5, p. 1559–1571, 1968

\bibitem{boyer} Robert H. Boyer \& Richard W. Lindquist, Maximal Analytic Extension of the Kerr
metric, Journal of Mathematical Physics, Vol. 8, p. 265, 1967.

\end{thebibliography}
\end{document}